\documentclass{article}

\usepackage{iclr2027_conference,times}
\usepackage{iftex}
\ifPDFTeX
  \usepackage[T1]{fontenc}
\fi
\usepackage{amsmath,amssymb,booktabs,array,graphicx,xcolor,tikz}
\usepackage{multirow,colortbl,placeins,wrapfig}
\usetikzlibrary{arrows.meta,positioning}
\usepackage[colorlinks=true,linkcolor=blue!45!black,citecolor=blue!45!black,urlcolor=blue!45!black]{hyperref}
\makeatletter\newcommand\captionof[1]{\def\@captype{#1}\caption}\makeatother
\makeatletter
\newcounter{algorithm}

\def\fps@algorithm{tbp}\def\ftype@algorithm{4}\def\ext@algorithm{loa}
\def\fnum@algorithm{Algorithm~\thealgorithm}
\newenvironment{algorithm}{\@float{algorithm}}{\end@float}
\makeatother
\newcounter{algline}
\newenvironment{algsteps}{\setcounter{algline}{0}\par\footnotesize\setlength{\parskip}{1.5pt}}{\par}
\newcommand{\algstep}[2]{\stepcounter{algline}\noindent\makebox[1.1em][r]{\scriptsize\arabic{algline}:}\hspace{0.5em}\hangindent=\dimexpr1.6em+#1em\relax\hangafter=1\hspace*{#1em}#2\par}
\newcommand{\algkw}[1]{\textbf{#1}}
\newcommand{\R}{\mathbb R}
\newcommand{\method}{scTAT}
\title{Rethinking One Gene, One Token:\newline Transcriptome-Aware Tokenization for\newline Single-Cell Foundation Models}
\iclrfinalcopy
\author{Jiaqi Xiong\textsuperscript{2}\thanks{Equal contribution.}, Jiaying Li\textsuperscript{3}\footnotemark[1], Yu Zheng\textsuperscript{4}, Xiang Zhou\textsuperscript{5}, Ziteng Wang\textsuperscript{6},\\
\textbf{Xinting Hu\textsuperscript{7}\footnotemark[2], Jiaxin Qi\textsuperscript{1}\thanks{Corresponding authors. Contact: \texttt{jxqi@cnic.cn}.}}\\[4pt]
\textsuperscript{1}CNIC\enspace
\textsuperscript{2}University of Oxford\enspace
\textsuperscript{3}Durham University\enspace
\textsuperscript{4}HNU\enspace
\textsuperscript{5}CUHK\enspace
\textsuperscript{6}UCSD\enspace
\textsuperscript{7}USTC}
\definecolor{rankpink}{HTML}{D64F8C}
\definecolor{rankblue}{HTML}{296CB7}
\newcommand{\rankone}[1]{\textbf{#1}}
\newcommand{\ranktwo}[1]{\underline{#1}}
\newcommand{\metricstd}[2]{\mbox{#1\,{\tiny$\pm$#2}}}

\begin{document}
\maketitle
\lhead{Preprint}

\begin{abstract}
Single-cell foundation models typically treat each gene as a token, which is both costly and wasteful. A cell can express thousands of genes, so one token per gene makes self-attention expensive, yet many of these genes are detected at low levels and carry too little signal to deserve a token of their own. Genes that are expressed together could instead share a token, leaving separate tokens for genes that stand out. We introduce \method{} (\textbf{T}ranscriptome-\textbf{A}ware \textbf{T}okenization), which pairs a gene-set vocabulary with a per-cell tokenizer. The vocabulary holds overlapping gene sets that recur with stable relative expression across the pretraining corpus. Using these sets raises two difficulties. First, which genes should share a token depends on the cell, whereas existing multi-gene tokens fix the grouping in advance. Second, gene sets of different sizes must be compared with single genes on a common scale. The tokenizer therefore decides per cell, letting gene sets and single genes compete for token positions under an expression score normalized by the square root of set size. Training still reconstructs individual genes, so predictions stay at gene resolution. In frozen-embedding cell-type annotation, \method{} matches the strongest one-gene-per-token model with half the token budget. Its advantage over one-gene-per-token models at the same budget grows as the budget shrinks and the tokenizer assigns more positions to gene sets. Encoding the same genes with gene-set tokens also nearly halves the Transformer's matrix-multiplication cost. These results show that a token need not hold exactly one gene: when the tokenizer decides per cell which genes share a token, a fixed budget covers more of each transcriptome at no cost in annotation accuracy.
\end{abstract}

\section{Introduction}
\begin{wrapfigure}[14]{r}{0.44\textwidth}
    \vspace{-\intextsep}
    \centering
    \includegraphics[width=\linewidth]{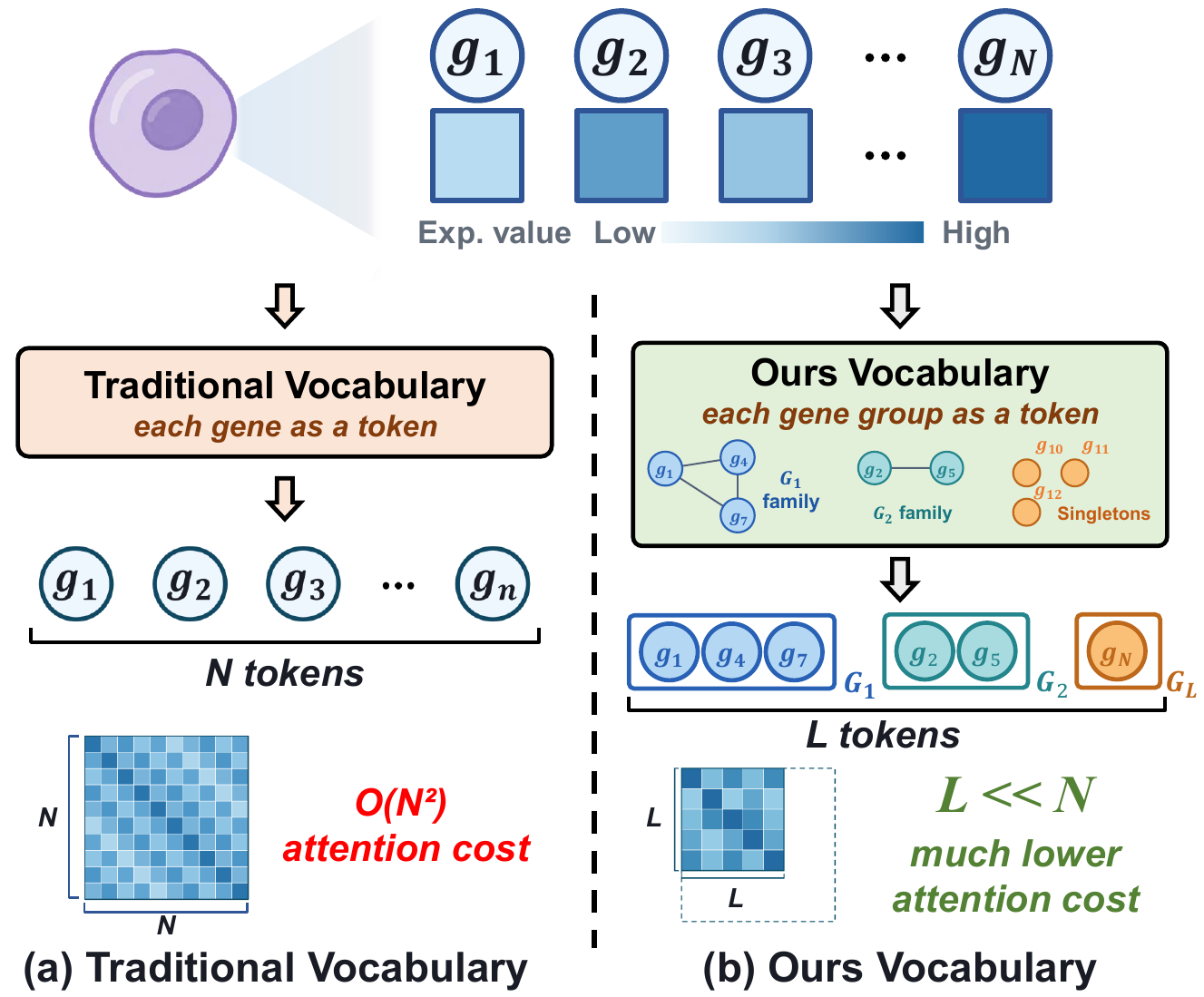}
    \vspace{-2.2em}
    \caption{(a) With one gene per token, $N$ genes take $N$ tokens and $O(N^2)$ attention. (b) With gene-set tokens, \method{} fits the same genes into $L\ll N$ tokens.}
    \label{fig:teaser}
\end{wrapfigure}
Single-cell foundation models are pretrained on tens of millions of cells~\citep{scgpt,scfoundation} and learn cell and gene representations that transfer to cell-type annotation, data integration, and perturbation analysis~\citep{scgpt,geneformer,uce}. Expression modeling also supports gene regulatory network inference~\citep{genie3,scprint,xatgrn}. Most single-cell foundation models adopt the recipe of language models: a cell is written as a sequence of gene tokens, and a Transformer learns from masked or ranked expression~\citep{scbert,geneformer,scgpt}. Under this recipe, the token determines both what the model sees of a cell and how much each cell costs to encode.

However, these two roles pull in opposite directions when each token holds one gene. A cell expresses thousands of genes, so showing the model more of them means a longer input, yet the cost of self-attention grows quadratically with input length~\citep{transformer}. Models that give each gene its own token, such as Geneformer and scGPT~\citep{geneformer,scgpt}, therefore keep only a subset of each cell's genes. Many of the remaining genes are detected at low levels and carry little signal on their own, yet each still occupies a full position. Expression-weighted sampling favors highly expressed genes when drawing this subset~\citep{scbenchmark,cocos}, but with one gene per position, the token budget still caps how many genes the model sees.

A direct way to lift this cap is to let one token carry several genes, so the same budget reaches more of each transcriptome. Several single-cell models take this route, grouping genes by curated pathways or regulons in TOSICA~\citep{tosica}, learned gene-module projections in Stack~\citep{stack}, or protein-informed soft assignment in scREBOUND~\citep{screbound}. Each grouping is fixed before any cell is seen, which causes two problems. First, relationships the grouping does not encode are lost at the input: genes outside curated pathways never enter a pathway token, and co-varying genes in different groups never share a token. Second, which genes belong together depends on the cell's type and state. Gene--gene co-expression differs across cell types~\citep{cscore}: cell-cycle genes rise together in dividing cells~\citep{tirosh2016}, and interferon-stimulated genes during viral infection~\citep{isg2014}. Conversely, some identities rest on a single highly expressed gene, such as insulin in beta cells. A fixed grouping spends positions on programs a cell does not run and blurs such markers.

To address these two problems, \textbf{we introduce \method{} (Transcriptome-Aware Tokenization)}, which pairs a corpus-mined vocabulary of overlapping gene sets with a per-cell tokenizer. The vocabulary is built along a gene-association graph, which keeps the combinatorial search tractable. It admits sets whose members co-occur often and keep stable relative levels, using a compression-based score inspired by the minimum description length (MDL) principle~\citep{rissanen1978,krimp}. The tokenizer lets gene sets and single genes compete for each cell's positions under a size-normalized expression score that balances coverage against per-gene signal. Reconstruction targets remain individual genes, so predictions stay at gene resolution.

The vocabulary and the tokenizer address these two problems in turn. Mined from co-occurrence in the pretraining corpus, the vocabulary can group co-varying genes whether or not they share a curated pathway, and because its sets overlap, no gene is locked into a single group. The tokenizer selects a gene set only when its members are jointly expressed in the cell, while a gene that dominates its set, such as insulin in a beta cell, keeps a token of its own; as positions become scarce, it assigns more of them to gene sets, which lifts the budget cap. In frozen-embedding annotation on ten datasets, the advantage of \method{} over one-gene-per-token models trained on the same corpus grows as the budget shrinks, its gene sets are enriched for known pathways and physical associations, and its frozen encoder remains competitive on gene-level tasks.

Our main contributions are as follows:
\begin{itemize}
\item We identify the token as a bottleneck: one gene per token caps how many genes a model sees, while existing multi-gene tokens fix their grouping in advance, losing relationships outside their groups and ignoring cell-dependent co-expression.
\item We construct a gene-set vocabulary and design a per-cell tokenizer: graph-constrained search and an MDL-inspired score admit overlapping gene sets that recur with stable relative expression, and the tokenizer lets these sets compete with single genes for each cell's token positions, so which genes share a token follows each cell's expression.
\item We conduct a comprehensive evaluation on ten cell-type annotation datasets and two gene-level tasks, with efficiency, ablation, and pathway analyses. \method{} matches the strongest one-gene-per-token model with half the token budget and nearly halves the Transformer FLOPs needed to encode the same genes.
\end{itemize}

\section{Related Work}

\paragraph{What a token holds in single-cell Transformers.}
Most single-cell foundation models give each token one gene. The token may combine gene identity with an expression value, as in scBERT~\citep{scbert} and scGPT~\citep{scgpt}, or encode expression through the gene's rank position, as in Geneformer~\citep{geneformer} and tGPT~\citep{tgpt}; scMulan adds metadata and task tokens~\citep{scmulan}. Biological knowledge typically enriches the gene embedding itself: GeneCompass injects regulatory and functional priors~\citep{genecompass}, and UCE uses protein-language-model embeddings~\citep{uce}. \method{} instead lets a gene-association graph restrict which genes may share a token and corpus statistics decide which sets are admitted.

\paragraph{Tokenizers in biological foundation models.}
Genomic models vary in their input representations: DNABERT uses fixed $k$-mers and DNABERT-2 learns byte-pair-encoding merges~\citep{dnabert,dnabert2}. MutBERT represents each nucleotide position as a probability distribution derived from population allele frequencies~\citep{long2025mutbert}. Expression profiles have no gene order, so single-cell models that group genes into tokens take membership from other sources. TOSICA uses curated pathways or regulons, masking a trainable projection so that each token reads a specified gene set~\citep{tosica}; Stack learns gene-module projections jointly with an encoder that attends across genes and cells~\citep{stack}; scREBOUND learns a protein-informed soft assignment that pools genes into meta-genes~\citep{screbound}. scConcept samples disjoint gene panels to form two views of each cell and learns cell representations through contrastive pretraining~\citep{scconcept}. Whereas TOSICA, Stack, and scREBOUND fix their grouping before any cell is seen, \method{} selects per cell which gene sets and single genes enter the input. Its gene sets are mined from the training corpus with a candidate-local variant of compression-based itemset mining~\citep{krimp} that adds expression-consistency and graph terms.

\paragraph{Efficient single-cell encoders.}
Other work reduces cost elsewhere in the model. Performer approximates softmax attention~\citep{performer}; xTrimoGene exploits expression sparsity with an asymmetric encoder--decoder~\citep{xtrimogene}; scBenchmark evaluates expression-weighted sampling and alternative encoders~\citep{scbenchmark}; CellPLM treats whole cells as tokens and attends across cells~\citep{cellplm}. In vision Transformers, token merging and pruning shorten the sequence inside the network by combining similar tokens or dropping uninformative ones~\citep{tome,dynamicvit}. \method{} instead decides before encoding which genes may share a token, from corpus statistics, so every token keeps an explicit gene membership. Shortening the gene sequence this way is complementary to efficient attention and to gene sampling.

\section{Method}
\label{sec:method}
\subsection{Overview}
\begin{figure}[t]
    \centering
    \includegraphics[width=\linewidth]{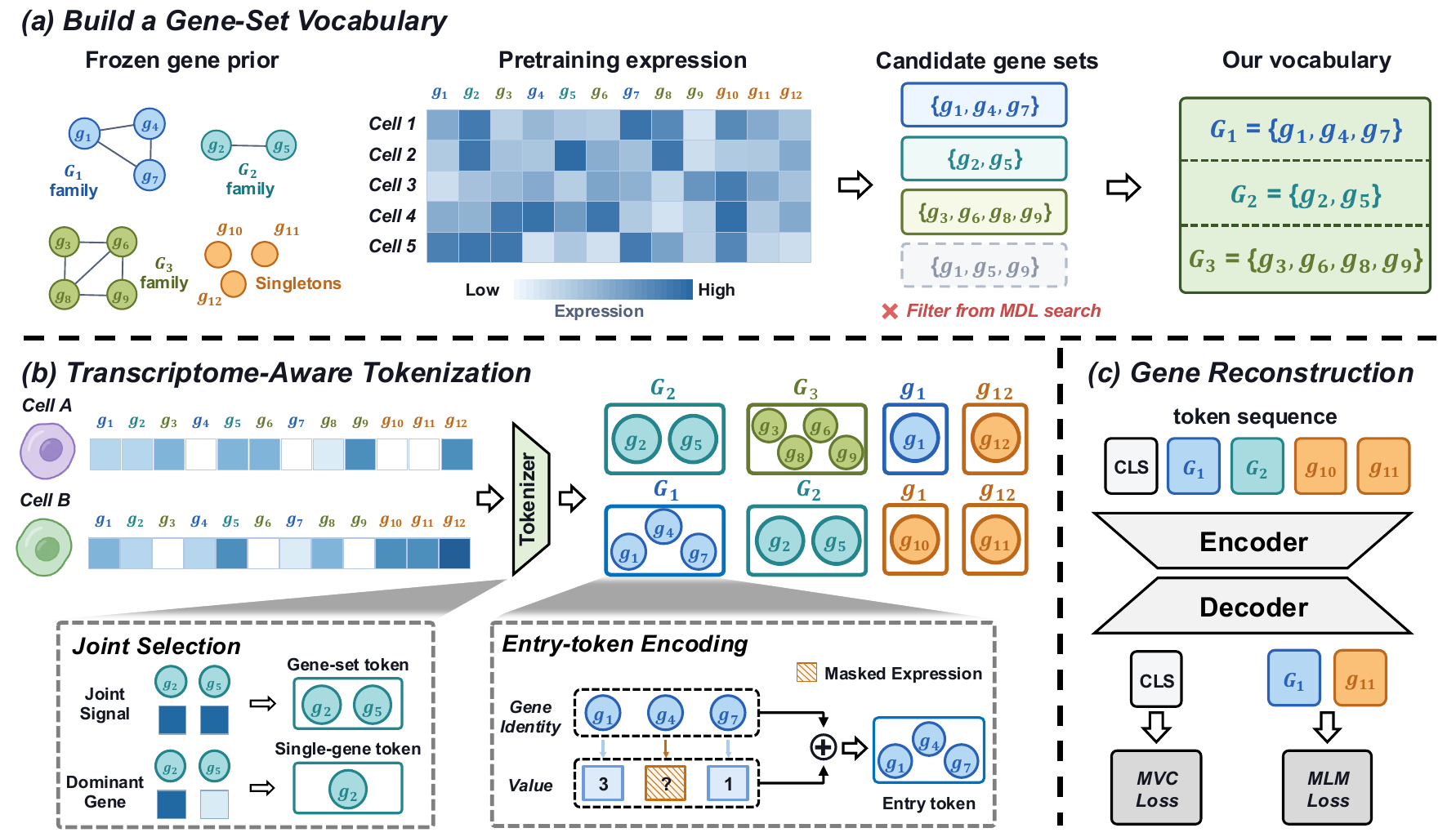}
    \caption{\textbf{Overview of \method{}.} \textbf{(a)} A gene graph proposes candidate gene sets, and an MDL-inspired score admits those that recur with stable relative expression. \textbf{(b)} For each cell, gene sets and single genes compete for token positions. \textbf{(c)} A Transformer encodes the tokens and reconstructs masked genes from the tokens (MLM) and from CLS (MVC).}
\label{fig:framework}
\end{figure}

\method{} builds a gene-set vocabulary offline (Section~\ref{sec:vocab}), selects gene sets and single genes for each cell (Section~\ref{sec:tokenize}), and trains a Transformer to reconstruct masked genes (Section~\ref{sec:recon}; Figure~\ref{fig:framework}).

Let $\mathcal G$ be the gene universe and $x\in\R_{\geq0}^{|\mathcal G|}$ a cell's processed expression profile; a zero means the gene was not detected. The active genes are $A(x)=\{g:x_g>0\}$. The vocabulary $\mathcal V$ contains two kinds of entries, each an explicit set of gene identities: \emph{gene sets} (multi-gene entries) and \emph{single genes} (singletons, one per gene in $\mathcal G$). Entries may overlap. An entry is shared across cells, whereas a token is the entry's instance in one cell, carrying that cell's expression. For each cell, tokenization selects at most $K$ entries $\mathcal T(x)\subseteq\mathcal V$ and adds a CLS token for the cell readout; when $K$ is small, some active genes remain unrepresented.

\subsection{Vocabulary construction}\label{sec:vocab}
\paragraph{Proposing candidates on a gene graph.}
The number of possible gene sets grows combinatorially, so we search only within neighborhoods of a gene graph. We build a mutual $k$-nearest-neighbor graph from the normalized gene embeddings of pretrained scGPT: two genes are connected when each ranks among the other's top-$k$ neighbors by cosine similarity, and the similarity is the edge weight. Its edge weights also enter the coherence term of the admission score below; only the graph comes from scGPT, and no scGPT weights initialize \method{}.

\paragraph{Admitting gene sets by compression.}
A gene set deserves a token when it recurs across many cells and its members keep similar relative expression, since one token then summarizes them with little loss. We measure both properties on a discovery sample of 50,000 training cells, split into two disjoint subsets that must each support the candidate; no evaluation cells are used. We combine them into a score following the minimum description length (MDL) principle~\citep{rissanen1978}, as in Krimp itemset mining~\citep{krimp}: a gene set is worth adding when encoding its members as one unit in the cells that contain all of them, plus the cost of describing the set, is shorter than encoding the members separately. For a candidate $S$, let $I_S\!=\!\{i:S\subseteq A(x_i)\}$ be its supporting cells and $n_S\!=\!|I_S|$; let $\ell_g$ and $\ell_S$ be the negative log smoothed occurrence frequencies of gene $g$ and of $S$, and let $L_{\mathrm{desc}}(S)$ be the cost of describing the size and members of $S$. The admission score is
\begin{equation}
\label{eq:gain}
F(S)=n_S\bigl(\textstyle\sum_{g\in S}\ell_g-\ell_S\bigr)-L_{\mathrm{desc}}(S)-n_S\mathcal R(S).
\end{equation}
The first term is the code length saved by encoding $S$ as one unit in its $n_S$ supporting cells; it is large when the members co-occur more often than their individual frequencies predict. The second term charges for adding $S$ to the vocabulary. The third term goes beyond a strict code length and penalizes unstable or weakly connected sets: each supporting cell pays $\mathcal R(S)=D(S)+(1-H(S))/2$, where $D(S)$ is the mean base-two Jensen--Shannon divergence of the cells' within-set expression proportions from their mean, and $H(S)$ is the mean edge weight of a maximum spanning tree on $S$. A set therefore scores highly when its members co-occur, keep stable proportions, and are tightly connected in the graph; Appendix~\ref{app:vocabulary-construction} gives the full definitions.

\paragraph{Searching for gene sets.}
Starting from each gene as an anchor $a$, we greedily extend a candidate with genes from the anchor's neighbor set $\mathcal N(a)$ while the admission score improves and the support constraint holds, so every candidate lies in $\{a\}\cup\mathcal N(a)$ (Algorithm~\ref{alg:vocab}). Adding all single genes to the admitted sets completes the vocabulary $\mathcal V$, which contains 79,536 gene sets of 2--32 genes (median 4) and 60,694 single genes (Appendix Figure~\ref{fig:vocabulary-statistics}).

\subsection{Transcriptome-aware tokenization}\label{sec:tokenize}
\paragraph{Selecting vocabulary entries.}
A gene set is eligible for a cell only when all of its members are active, $S\subseteq A(x)$; every active gene is also eligible as a single gene. Let $C$ be the union of genes already represented, initially empty. An eligible entry would add the expression
\begin{equation}
\label{eq:selection}
\Delta_x(S\mid C)=\sum_{g\in S\setminus C}x_g
\end{equation}
carried by its $n_C(S)=|S\setminus C|$ newly represented genes. \method{} places gene sets and single genes in one candidate pool and scores each by
\begin{equation}
\label{eq:joint}
s(S\mid C)=\frac{\Delta_x(S\mid C)}{\sqrt{n_C(S)}} .
\end{equation}
At each step, the tokenizer selects the highest-scoring entry, adds its genes to $C$, and recomputes scores; entries adding no new gene are skipped, and ties favor single genes. Selection stops when the budget is filled or every active gene is represented. This rule is the \method{} tokenizer.

\paragraph{Why the square root.}
Normalizing by $\sqrt{n_C(S)}$ balances coverage against per-gene signal. Without normalization, the new expression $\Delta_x$ grows with every added member, so the score favors large gene sets. Dividing by $n_C(S)$ instead scores the mean expression of the new members, which never exceeds the score of the most highly expressed member as a single gene, so no gene set would ever be selected. The square root lies between these extremes: a gene set wins when its new members are jointly well expressed, and a single gene wins when it dominates, so the mix of gene sets and single genes depends on each cell's profile. For example, two new genes with expression 3 each score $6/\sqrt{2}\approx4.2$ as a set and beat either gene alone, whereas a gene with expression 10 and a partner with expression 1 score $11/\sqrt{2}\approx7.8$ as a set and lose to the first gene alone.

\paragraph{Encoding token instances.}
Each selected entry $S$ becomes one token with features
\begin{equation}
t_S=\operatorname{Normalize}\!\left(b_S+\phi(v_S)+c_S\right),\qquad b_S=\frac1{|S|}\sum_{g\in S}e_g,\qquad v_S=\frac1{|V_S|}\sum_{g\in V_S}x_g,
\end{equation}
where $e_g$ is a trainable gene embedding, $V_S$ the members whose values are not masked, $\phi$ a bounded scalar-expression encoder, and $c_S$ a learned embedding of the masked-member fraction and entry size. Masking is shared across all occurrences of a gene, and an entry with no visible members receives a mask indicator in place of $\phi(v_S)$. A Transformer, trained from scratch, encodes the tokens together with CLS into token outputs $h_S$ and a cell embedding $z$.

\subsection{Reconstruction in the original gene space}\label{sec:recon}
Although inputs may bundle several genes into one token, every prediction target is a single gene. Let $M$ be the masked active genes and $J_g=\{S\in\mathcal T(x):g\in S\}$ the selected entries containing gene $g$. The covered targets are $M_C=\{g\in M:J_g\neq\varnothing\}$. For $g\in M_C$, we form a reconstruction context from the corresponding token outputs:
\begin{equation}
r_g=|J_g|^{-1}\sum_{S\in J_g}Wh_S.
\end{equation}
Here $W$ is a learned projection. A token-based decoder combines $r_g$ with a learned gene query to predict the masked value, yielding the masked-expression objective (MLM) on $M_C$. A cell-based decoder uses $z$ and a gene-specific query to predict every target in $M$, yielding the masked-value reconstruction objective (MVC). Uncovered masked genes contribute only to MVC. Both squared-error objectives are optimized jointly; downstream evaluation uses the frozen CLS embedding.

\paragraph{Pretraining.}
We pretrain on 1.8 million human cells, with 13,780 cells held out for validation, from the scBenchmark corpus~\citep{scbenchmark}, which assembles 54.6 million CELLxGENE cells following scGPT~\citep{scgpt}. The main \method{} models are trained from scratch for 10 epochs at 64, 128, 256, and 512 tokens with the \method{} tokenizer; Appendix Table~\ref{tab:config} lists all settings.

\section{Experiments}
\subsection{Experimental Setup}
\label{sec:implementation}
\paragraph{Datasets and splits.}
Cell-type annotation follows scBenchmark~\citep{scbenchmark}: frozen CLS embeddings are evaluated by KNN and a linear probe on its ten datasets (Multiple Sclerosis, MS; Myeloid, Myel.; Pancreas, Panc.; Lupus; Dengue, Deng.; irColitis, irCol.; Leptomeningeal, Lepto.; Myasthenia, Myas.; Scanorama, Scan.; and scFoundation, scF.), none of them in the pretraining corpus, and we report macro accuracy averaged over five cell-level splits (Appendix~\ref{app:annotation-variability}). Gene-network ranking follows the Han hESC task of scPRINT~\citep{scprint} (Appendix~\ref{app:grn-evaluation}), and perturbation prediction the Norman\_1 task of PertEval-scFM~\citep{perteval} (Appendix~\ref{app:perturbation-evaluation}).

\paragraph{Baselines.}
We compare with scGPT~\citep{scgpt}, Geneformer V2~\citep{geneformer}, the Transformer released with scBenchmark (hereafter scBenchmark)~\citep{scbenchmark}, and CoCoS~\citep{cocos}. scBenchmark and CoCoS share the pretraining data, the number of epochs, and the six-layer, 256-dimensional Transformer backbone of \method{}. Each baseline receives up to 512 genes, sampled without replacement uniformly (Uniform) or in proportion to expression (Weighted); TAT denotes the \method{} tokenizer. To isolate the effect of the token, we also pretrain the \method{} architecture with single-gene tokens only at different budget, sampling expressed genes in proportion to expression as in weighted scBenchmark.

\begin{table}[t]
\centering
\caption{Frozen KNN annotation accuracy (\%), averaged over five splits. \rankone{Bold} and \ranktwo{underlined} values are the best and second-best mean in each column.}\label{tab:main-knn}
\begingroup\setlength{\tabcolsep}{1.3pt}\renewcommand{\arraystretch}{1.12}\footnotesize
\begin{tabular*}{\linewidth}{@{\extracolsep{\fill}}llc*{10}{c}c@{}}\toprule
\textbf{Model} & \textbf{Input} & \textbf{Budget} & \textbf{MS} & \textbf{Myel.} & \textbf{Panc.} & \textbf{Lupus} & \textbf{Deng.} & \textbf{irCol.} & \textbf{Lepto.} & \textbf{Myas.} & \textbf{Scan.} & \textbf{scF.} & \textbf{Avg} \\
\midrule
\multirow{2}{*}{scBenchmark} & Uniform & \multirow{2}{*}{512} & \ranktwo{85.47} & 70.35 & 96.56 & 64.93 & 82.99 & 53.27 & 59.54 & 80.78 & 92.34 & 64.46 & 75.07 \\
 & \cellcolor{black!4}Weighted &  & \cellcolor{black!4}\rankone{85.61} & \cellcolor{black!4}71.70 & \cellcolor{black!4}97.00 & \cellcolor{black!4}64.93 & \cellcolor{black!4}\rankone{85.15} & \cellcolor{black!4}\rankone{59.16} & \cellcolor{black!4}90.81 & \cellcolor{black!4}\ranktwo{84.33} & \cellcolor{black!4}94.80 & \cellcolor{black!4}64.81 & \cellcolor{black!4}79.83 \\
\midrule[0.3pt]
\multirow{2}{*}{CoCoS} & Uniform & \multirow{2}{*}{512} & 81.91 & 66.69 & 92.61 & \rankone{72.55} & 82.68 & 52.48 & 61.47 & 80.63 & 90.40 & \ranktwo{66.48} & 74.79 \\
 & \cellcolor{black!4}Weighted &  & \cellcolor{black!4}82.15 & \cellcolor{black!4}67.77 & \cellcolor{black!4}93.04 & \cellcolor{black!4}\rankone{72.55} & \cellcolor{black!4}84.53 & \cellcolor{black!4}56.18 & \cellcolor{black!4}89.37 & \cellcolor{black!4}83.65 & \cellcolor{black!4}93.34 & \cellcolor{black!4}\rankone{66.83} & \cellcolor{black!4}78.94 \\
\midrule[0.3pt]
\multirow{2}{*}{scGPT} & Uniform & \multirow{2}{*}{512} & 75.42 & 67.66 & 91.96 & 66.49 & 81.05 & 49.30 & 44.67 & 78.34 & 90.30 & 62.99 & 70.82 \\
 & \cellcolor{black!4}Weighted &  & \cellcolor{black!4}75.46 & \cellcolor{black!4}68.82 & \cellcolor{black!4}92.71 & \cellcolor{black!4}66.49 & \cellcolor{black!4}84.23 & \cellcolor{black!4}55.51 & \cellcolor{black!4}87.13 & \cellcolor{black!4}83.41 & \cellcolor{black!4}93.71 & \cellcolor{black!4}63.39 & \cellcolor{black!4}77.09 \\
\midrule[0.3pt]
\multirow{2}{*}{Geneformer V2} & Uniform & \multirow{2}{*}{512} & 70.71 & 62.90 & 89.49 & \ranktwo{71.98} & 76.98 & 46.58 & 44.37 & 74.31 & 79.66 & 59.40 & 67.64 \\
 & \cellcolor{black!4}Weighted &  & \cellcolor{black!4}71.28 & \cellcolor{black!4}65.35 & \cellcolor{black!4}90.16 & \cellcolor{black!4}\ranktwo{71.98} & \cellcolor{black!4}78.82 & \cellcolor{black!4}50.91 & \cellcolor{black!4}72.37 & \cellcolor{black!4}78.28 & \cellcolor{black!4}87.12 & \cellcolor{black!4}60.76 & \cellcolor{black!4}72.70 \\
\midrule[0.3pt]
\multirow{4}{*}{\method{}} & \multirow{4}{*}{TAT} & 64 & 71.89 & 69.56 & \ranktwo{97.57} & 66.96 & 83.49 & 55.22 & 92.23 & 82.46 & 93.72 & 62.97 & 77.61 \\
 &  & 128 & 76.55 & 71.86 & \rankone{97.68} & 68.01 & 84.17 & 55.86 & 93.04 & 83.47 & 94.36 & 64.86 & 78.99 \\
 &  & 256 & 80.63 & \rankone{72.80} & 97.23 & 68.37 & \ranktwo{84.93} & 56.76 & \rankone{93.27} & 84.32 & \ranktwo{94.97} & 65.89 & \ranktwo{79.92} \\
 &  & \cellcolor{black!4}512 & \cellcolor{black!4}81.71 & \cellcolor{black!4}\ranktwo{72.75} & \cellcolor{black!4}97.52 & \cellcolor{black!4}68.80 & \cellcolor{black!4}84.90 & \cellcolor{black!4}\ranktwo{57.23} & \cellcolor{black!4}\ranktwo{93.09} & \cellcolor{black!4}\rankone{84.37} & \cellcolor{black!4}\rankone{95.25} & \cellcolor{black!4}65.52 & \cellcolor{black!4}\rankone{80.12} \\
\bottomrule\end{tabular*}\endgroup
\end{table}

\begin{table}[t]
\centering
\caption{Frozen linear-probe annotation accuracy (\%); settings as in Table~\ref{tab:main-knn}.}\label{tab:main-linear}
\begingroup\setlength{\tabcolsep}{1.3pt}\renewcommand{\arraystretch}{1.12}\footnotesize
\begin{tabular*}{\linewidth}{@{\extracolsep{\fill}}llc*{10}{c}c@{}}\toprule
\textbf{Model} & \textbf{Input} & \textbf{Budget} & \textbf{MS} & \textbf{Myel.} & \textbf{Panc.} & \textbf{Lupus} & \textbf{Deng.} & \textbf{irCol.} & \textbf{Lepto.} & \textbf{Myas.} & \textbf{Scan.} & \textbf{scF.} & \textbf{Avg} \\
\midrule
\multirow{2}{*}{scBenchmark} & Uniform & \multirow{2}{*}{512} & \rankone{87.18} & 69.74 & 96.07 & 76.67 & 82.44 & 52.29 & 66.47 & 81.96 & 92.98 & 67.31 & 77.31 \\
 & \cellcolor{black!4}Weighted &  & \cellcolor{black!4}\ranktwo{87.09} & \cellcolor{black!4}70.58 & \cellcolor{black!4}96.74 & \cellcolor{black!4}76.67 & \cellcolor{black!4}84.70 & \cellcolor{black!4}56.16 & \cellcolor{black!4}91.63 & \cellcolor{black!4}85.00 & \cellcolor{black!4}94.60 & \cellcolor{black!4}67.55 & \cellcolor{black!4}81.07 \\
\midrule[0.3pt]
\multirow{2}{*}{CoCoS} & Uniform & \multirow{2}{*}{512} & 86.12 & 68.28 & 95.00 & 78.03 & 83.63 & 53.02 & 68.96 & 81.96 & 92.40 & \rankone{69.38} & 77.68 \\
 & \cellcolor{black!4}Weighted &  & \cellcolor{black!4}86.10 & \cellcolor{black!4}68.65 & \cellcolor{black!4}95.45 & \cellcolor{black!4}78.03 & \cellcolor{black!4}\ranktwo{85.53} & \cellcolor{black!4}\ranktwo{56.22} & \cellcolor{black!4}91.13 & \cellcolor{black!4}84.82 & \cellcolor{black!4}94.27 & \cellcolor{black!4}\ranktwo{69.29} & \cellcolor{black!4}80.95 \\
\midrule[0.3pt]
\multirow{2}{*}{scGPT} & Uniform & \multirow{2}{*}{512} & 81.71 & 70.28 & 93.78 & 76.66 & 83.80 & 52.39 & 56.43 & 81.92 & 92.38 & 68.52 & 75.79 \\
 & \cellcolor{black!4}Weighted &  & \cellcolor{black!4}81.55 & \cellcolor{black!4}71.24 & \cellcolor{black!4}94.39 & \cellcolor{black!4}76.66 & \cellcolor{black!4}\rankone{86.13} & \cellcolor{black!4}\rankone{57.33} & \cellcolor{black!4}90.46 & \cellcolor{black!4}\rankone{85.60} & \cellcolor{black!4}\ranktwo{95.23} & \cellcolor{black!4}68.52 & \cellcolor{black!4}80.71 \\
\midrule[0.3pt]
\multirow{2}{*}{Geneformer V2} & Uniform & \multirow{2}{*}{512} & 80.81 & 65.90 & 92.87 & \rankone{80.02} & 80.30 & 50.72 & 50.32 & 79.95 & 90.92 & 63.79 & 73.56 \\
 & \cellcolor{black!4}Weighted &  & \cellcolor{black!4}81.29 & \cellcolor{black!4}67.03 & \cellcolor{black!4}94.00 & \cellcolor{black!4}\rankone{80.02} & \cellcolor{black!4}81.65 & \cellcolor{black!4}54.04 & \cellcolor{black!4}83.12 & \cellcolor{black!4}82.65 & \cellcolor{black!4}93.98 & \cellcolor{black!4}64.16 & \cellcolor{black!4}78.19 \\
\midrule[0.3pt]
\multirow{4}{*}{\method{}} & \multirow{4}{*}{TAT} & 64 & 78.76 & 69.47 & \rankone{97.54} & 76.64 & 83.00 & 54.20 & 92.50 & 83.94 & 94.23 & 66.93 & 79.72 \\
 &  & 128 & 81.67 & 70.82 & 97.33 & 77.08 & 84.07 & 54.54 & 92.87 & 84.77 & 94.62 & 67.50 & 80.53 \\
 &  & 256 & 85.13 & \rankone{71.56} & 96.96 & 77.59 & 84.43 & 54.84 & \ranktwo{93.07} & 85.25 & 94.93 & 68.37 & \ranktwo{81.21} \\
 &  & \cellcolor{black!4}512 & \cellcolor{black!4}85.74 & \cellcolor{black!4}\ranktwo{71.31} & \cellcolor{black!4}\ranktwo{97.43} & \cellcolor{black!4}\ranktwo{78.11} & \cellcolor{black!4}84.18 & \cellcolor{black!4}55.28 & \cellcolor{black!4}\rankone{93.25} & \cellcolor{black!4}\ranktwo{85.26} & \cellcolor{black!4}\rankone{95.28} & \cellcolor{black!4}68.65 & \cellcolor{black!4}\rankone{81.45} \\
\bottomrule\end{tabular*}\endgroup
\end{table}

\subsection{Results and Analysis}
\begin{figure}[!t]
\centering
\includegraphics[width=\linewidth]{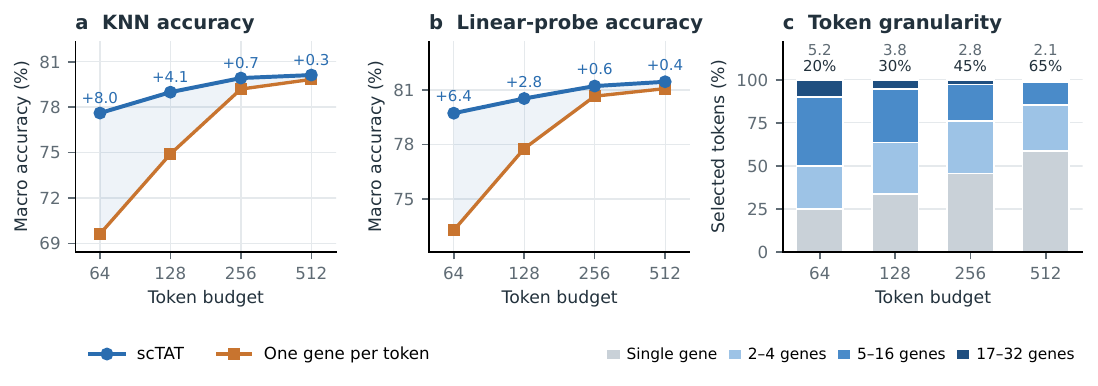}
\caption{\method{} across token budgets. \textbf{a,b}, Annotation accuracy of \method{} and of models with one gene per token; numbers give the margin of \method{}. \textbf{c}, Selected tokens by gene-set size on validation cells; numbers above the bars give mean coverage of expressed genes and covered genes per token.}
\label{fig:token-budget}
\end{figure}

\paragraph{Q1. How does annotation accuracy change with the token budget?}
Tables~\ref{tab:main-knn} and~\ref{tab:main-linear} compare frozen embeddings from \method{} at 64 to 512 tokens with four pretrained baselines, each given 512 genes by uniform or expression-weighted sampling. With 256 tokens, half the baselines' budget, \method{} matches the strongest baseline, expression-weighted scBenchmark (79.92\% versus 79.83\% KNN and 81.21\% versus 81.07\% linear-probe accuracy), and with 128 or 64 tokens it remains above every uniformly sampled 512-token baseline. Against models pretrained with one gene per token at the same budget (Figure~\ref{fig:token-budget}a,b), its margin widens from 0.3 to 8.0 KNN points and from 0.4 to 6.4 linear-probe points as the budget falls from 512 to 64 tokens, so gene-set tokens matter most when positions are scarce. At 512 tokens, \method{} is also the highest under both probes, 0.29 and 0.38 points above expression-weighted scBenchmark (paired over five splits: $0.29\pm0.14$ and $0.38\pm0.20$; Appendix~\ref{app:annotation-variability}).

\paragraph{Q2. How does token granularity change with the token budget?}
Because gene sets and single genes compete for positions, the tokenizer's granularity is an outcome rather than a fixed design (Figure~\ref{fig:token-budget}c). On validation cells, gene sets fill 74.9\%, 66.2\%, 54.4\%, and 41.3\% of the selected positions at 64, 128, 256, and 512 tokens, each token covers 5.2 to 2.1 genes on average, and the selected tokens reach 20\% to 65\% of a cell's expressed genes. At 64 tokens, half of the positions hold sets of at least five genes, against 15\% at 512 tokens. About two-thirds of the first 128 positions go to jointly expressed gene sets, but only about one in three of the positions added between 128 and 512 tokens; because the greedy order does not depend on the budget, the sets chosen early are kept.

\paragraph{Q3. How much encoder computation do gene-set tokens save?}
If a token carries several genes, fewer tokens should suffice to represent a given set of genes. We test this with the main models at 64, 128, 256, and 512 tokens: each cell's selected gene union is expanded into single-gene inputs for a fixed scBenchmark checkpoint, and both encoders process the same 10,000 cells (Table~\ref{tab:efficiency}).

\begin{table}[t]
\centering
\begin{minipage}[t]{0.53\linewidth}\vspace{0pt}\centering
\caption{Cost of encoding the same genes on 10,000 cells, per cell except peak memory (GiB, per batch of 64); bold marks the lower cost.}\label{tab:efficiency}

\begingroup\footnotesize\setlength{\tabcolsep}{1.8pt}\renewcommand{\arraystretch}{1.05}
\begin{tabular*}{\linewidth}{@{\extracolsep{\fill}}llrrrr@{}}
\toprule
\textbf{\method{} budget} & \textbf{Tokens} & \textbf{Len.}$\downarrow$ & \textbf{GFLOPs}$\downarrow$ & \textbf{GiB}$\downarrow$ & \textbf{ms}$\downarrow$ \\
\midrule
\multirow{2}{*}{64} & Single gene & 211 & 1.35 & 0.43 & 0.26 \\
 & \method{} & \textbf{65} & \textbf{0.33} & \textbf{0.20} & \textbf{0.17} \\
\midrule[0.3pt]
\multirow{2}{*}{128} & Single gene & 302 & 2.14 & 0.58 & 0.40 \\
 & \method{} & \textbf{124} & \textbf{0.68} & \textbf{0.22} & \textbf{0.30} \\
\midrule[0.3pt]
\multirow{2}{*}{256} & Single gene & 434 & 3.53 & 0.81 & 0.64 \\
 & \method{} & \textbf{231} & \textbf{1.44} & \textbf{0.29} & \textbf{0.53} \\
\midrule[0.3pt]
\multirow{2}{*}{512} & Single gene & 612 & 5.97 & 1.20 & 1.12 \\
 & \method{} & \textbf{406} & \textbf{3.08} & \textbf{0.43} & \textbf{1.00} \\
\bottomrule
\end{tabular*}
\endgroup

\end{minipage}\hfill
\begin{minipage}[t]{0.43\linewidth}\vspace{0pt}\centering
\caption{Ablations at 512 tokens (macro accuracy, \%).}\label{tab:ablation-selection}\label{tab:ablation-vocab}
\begingroup\setlength{\tabcolsep}{2.5pt}\renewcommand{\arraystretch}{1.08}\footnotesize
\renewcommand{\metricstd}[2]{\mbox{#1\,{\fontsize{6}{7}\selectfont$\pm$#2}}}
\begin{tabular*}{\linewidth}{@{\extracolsep{\fill}}lcc@{}}
\toprule
\textbf{Variant} & \textbf{KNN} & \textbf{Linear} \\
\midrule
\multicolumn{3}{@{}l}{\textit{(a) Gene sets first, ranked by}} \\
New expression & \metricstd{79.62}{0.12} & \metricstd{81.25}{0.16} \\
New gene count & \metricstd{79.37}{0.16} & \metricstd{80.92}{0.20} \\
Set size & \metricstd{78.65}{0.12} & \metricstd{80.55}{0.21} \\
Admission score & \metricstd{78.11}{0.18} & \metricstd{80.02}{0.18} \\
\midrule[0.3pt]
\multicolumn{3}{@{}l}{\textit{(b) Vocabulary}} \\
w/o MDL & \metricstd{79.43}{0.15} & \metricstd{80.67}{0.15} \\
w/o gene content & \metricstd{78.86}{0.12} & \metricstd{80.64}{0.22} \\
\midrule
\cellcolor{black!6}\method{} (ours) & \cellcolor{black!6}\metricstd{\textbf{80.12}}{0.11} & \cellcolor{black!6}\metricstd{\textbf{81.45}}{0.16} \\
\bottomrule
\end{tabular*}
\endgroup
\end{minipage}
\end{table}

\begin{figure}[!t]
\centering
\includegraphics[width=\linewidth]{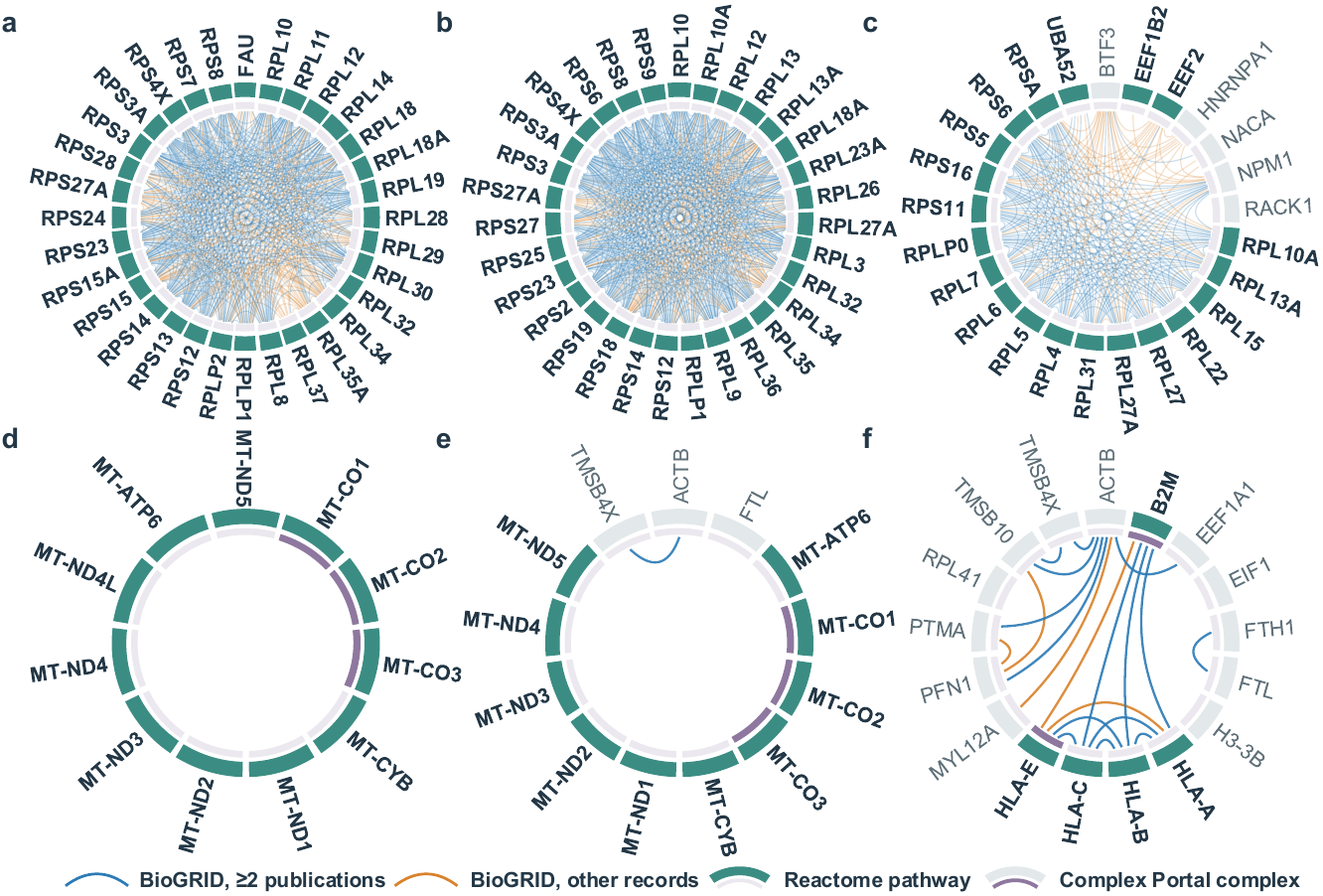}
\caption[Six frequently selected gene sets]{
Six frequently selected \method{} gene sets: \textbf{a,b}, peptide chain elongation; \textbf{c}, eukaryotic translation elongation; \textbf{d,e}, mitochondrial RNA processing and stability; \textbf{f}, endosomal/vacuolar pathway. Rings mark Reactome pathway (outer) and Complex Portal complex (inner) membership, and chords mark BioGRID physical associations~\citep{reactome,complexportal,biogrid}.

}\label{fig:biology-cases}
\end{figure}

At 512 tokens, encoding the selected genes as gene-set tokens rather than single-gene tokens lowers Transformer matrix-multiplication cost by 48.5\% and peak GPU memory by 63.8\%; at 64 tokens, it lowers them by 75.5\% and 53.9\%, and GPU encoding time by 34.5\%. The FLOPs savings grow as the budget shrinks because each token then covers more genes (Figure~\ref{fig:token-budget}c). Fewer tokens also mean less computation for the same accuracy: at full budget, \method{} at 256 tokens matches one gene per token at 512 tokens (Figure~\ref{fig:token-budget}a,b) with 40\% of the GFLOPs per cell (1.6 versus 4.0), whereas scGPT and Geneformer need 26 and 97~GFLOPs per cell at 512 tokens.

\paragraph{Q4. Which design choices matter?}
\textit{Selection.} Table~\ref{tab:ablation-selection}a compares the \method{} tokenizer, which scores gene sets and single genes in one pool, with four rules that rank gene sets by a fixed priority, select them before any single gene, and fill the remaining positions by expression-weighted sampling; each rule is used for both pretraining and evaluation at 512 tokens. The \method{} tokenizer is best, 0.50 KNN and 0.20 linear-probe points above the strongest alternative, ranking by new expression (the Expression-first rule), and the gap grows to 1.97 and 1.35 points at 128 tokens. Among the gene-set-first rules, ranking by newly covered expression (Equation~\ref{eq:selection}) beats ranking by newly covered gene count or set size. Ranking by the offline admission score does worst, which suggests that how reusable a set is across the corpus says little about whether it is informative in a given cell.

\textit{Vocabulary.} Two control vocabularies change only which gene sets are admitted (Table~\ref{tab:ablation-vocab}b; Appendix~\ref{app:vocabulary-construction}). \textit{w/o MDL} keeps the graph, the support requirement, and nearly the same size distribution but admits the most graph-coherent sets instead of scoring them; the admission score adds 0.69 KNN and 0.78 linear-probe points. \textit{w/o gene content} permutes gene labels within detection-frequency strata, preserving the number, sizes, and overlaps of entries. Under the same Expression-first selection, its tokens cover 1.77 instead of 2.24 genes each, coverage falls from 66.5\% to 55.4\% because a set enters only when all its members are expressed and permuted sets less often are, and accuracy drops by 0.76 and 0.61 points.

\paragraph{Supplementary gene-level tasks.}
We also test two gene-level uses of the frozen encoder (Table~\ref{tab:supplementary}). In Han hESC network ranking (a), \method{} ranks regulatory edges at 1.17 times the random baseline against the Full reference and 1.27 times against KO, second-best on Full and best on KO. In Norman\_1 perturbation prediction (b), \method{} performs close to scGPT and Geneformer, and the same holds on three further PertEval-scFM tasks (Appendix Table~\ref{tab:perturbation-additional}).

\begin{table}[!t]
\centering
\caption{Gene-level evaluation with frozen encoders. \textbf{(a)}~Han hESC network AUPRC against three references, without head selection; Random is the positive-edge prevalence, and \rankone{bold} and \ranktwo{underlined} values mark the best and second-best model. \textbf{(b)}~Norman single-gene perturbation prediction, area under the test-MSE curve for $s\in[0,0.7]$ (mean $\pm$ SD over three repetitions).}\label{tab:supplementary}\label{tab:grn}\label{tab:perturbation}
\begingroup
\fontsize{8}{9.8}\selectfont
\renewcommand{\arraystretch}{1.1}
\renewcommand{\metricstd}[2]{\mbox{#1\,{\fontsize{6}{7}\selectfont$\pm$#2}}}
\begin{minipage}[t]{0.54\linewidth}\centering
\textbf{(a) Network ranking (Han hESC)}\par\smallskip
\setlength{\tabcolsep}{3pt}
\begin{tabular*}{\linewidth}{@{\extracolsep{\fill}}lccc@{}}
\toprule
\textbf{Model} & \textbf{Full}$\uparrow$ & \textbf{ChIP}$\uparrow$ & \textbf{KO}$\uparrow$ \\
\midrule
\textit{Random} & \textit{0.0274} & \textit{0.2536} & \textit{0.0761} \\
\midrule[0.3pt]
GENIE3~\citep{genie3} & 0.0295 & 0.2783 & 0.0708 \\
scGPT~\citep{scgpt} & 0.0227 & 0.2360 & 0.0675 \\
Geneformer~\citep{geneformer} & \rankone{0.0370} & \rankone{0.3188} & 0.0619 \\
scPRINT~\citep{scprint} & 0.0286 & \ranktwo{0.2865} & \ranktwo{0.0728} \\
\midrule[0.3pt]
\method{} & \ranktwo{0.0319} & 0.2257 & \rankone{0.0965} \\
\bottomrule
\end{tabular*}
\end{minipage}\hfill
\begin{minipage}[t]{0.43\linewidth}\centering
\textbf{(b) Perturbation prediction (Norman)}\par\smallskip
\setlength{\tabcolsep}{3pt}
\begin{tabular*}{\linewidth}{@{\extracolsep{\fill}}lc@{}}
\toprule
\textbf{Model} & \textbf{MSE--$s$ area} ($\times10^{-2}$)$\downarrow$ \\
\midrule
Mean & \metricstd{4.705}{0.071} \\
Raw-expr.\ MLP & \metricstd{4.683}{0.050} \\
Geneformer (6L) & \metricstd{4.649}{0.053} \\
scFoundation & \metricstd{4.800}{0.099} \\
scGPT & \metricstd{4.650}{0.053} \\
\midrule[0.3pt]
\method{} & \metricstd{4.650}{0.055} \\
\bottomrule
\end{tabular*}
\end{minipage}
\endgroup
\end{table}

\begingroup
\renewcommand{\floatpagefraction}{0.9}

\subsection{Biological support for the vocabulary}\label{sec:biology-setup}\label{sec:biology-results}
\paragraph{External biological support.}
The admitted gene sets are enriched for known pathways and physical associations. On a prespecified panel of 1,000 gene sets, 20.10\% have a significant Reactome pathway hit~\citep{reactome}, against 3.87--4.83\% for stratified gene-label permutations; BioGRID physical associations~\citep{biogrid} show the same pattern. Random sets from the same graph neighborhoods reach 17.05\%, and corpus-based admission adds a further 1.18-fold for pathways and 1.38-fold for physical associations (Appendix~\ref{app:biology-details}). Figure~\ref{fig:biology-cases} shows six frequently selected gene sets: three translation programs whose members almost all belong to the named Reactome pathways, two mitochondrial RNA-processing sets that share complex~IV subunits, and an endosomal set containing the HLA-E/B2M complex. These sets group co-regulated or co-assembled genes: ribosomal proteins, mitochondrial transcripts processed by FASTK-family proteins~\citep{ohkubo2021}, and MHC class~I chains with B2M. Appendix~\ref{app:biology-details} describes each set.

\FloatBarrier
\endgroup

\section{Conclusion}
\method{} replaces the one-gene token with a choice made for each cell. A vocabulary mined from the pretraining corpus supplies overlapping gene sets whose members co-occur with stable relative expression, and a size-normalized expression score lets these sets compete with single genes for every position, while reconstruction targets remain individual genes. With frozen embeddings, \method{} matches the strongest one-gene-per-token model with half the tokens and exceeds one-gene-per-token models at every budget from 64 to 512 tokens, by the widest margin when tokens are scarce and the tokenizer assigns most positions to gene sets. It encodes the same genes with nearly half the Transformer FLOPs and less GPU memory, its gene sets are enriched for known pathways and physical associations, and its frozen encoder remains competitive on gene-level tasks.

These results show that what a token holds is a design choice rather than a fixed convention. Deciding per cell which genes share a token lets a fixed budget cover more of each transcriptome, with the largest benefit under tight budgets, where one gene per token must leave most of each transcriptome out. Because it shortens the input before encoding, this choice is complementary to efficient attention and to gene sampling. Natural next steps are combinatorial perturbation prediction as in GEARS~\citep{gears}, refining the vocabulary jointly with encoder training, and vocabularies for other species and modalities, such as chromatin accessibility or spatial transcriptomics.

\clearpage
\appendix

\section{Vocabulary analyses}\label{app:biology-details}
\begin{figure}[!htbp]
\centering
\includegraphics[width=\linewidth]{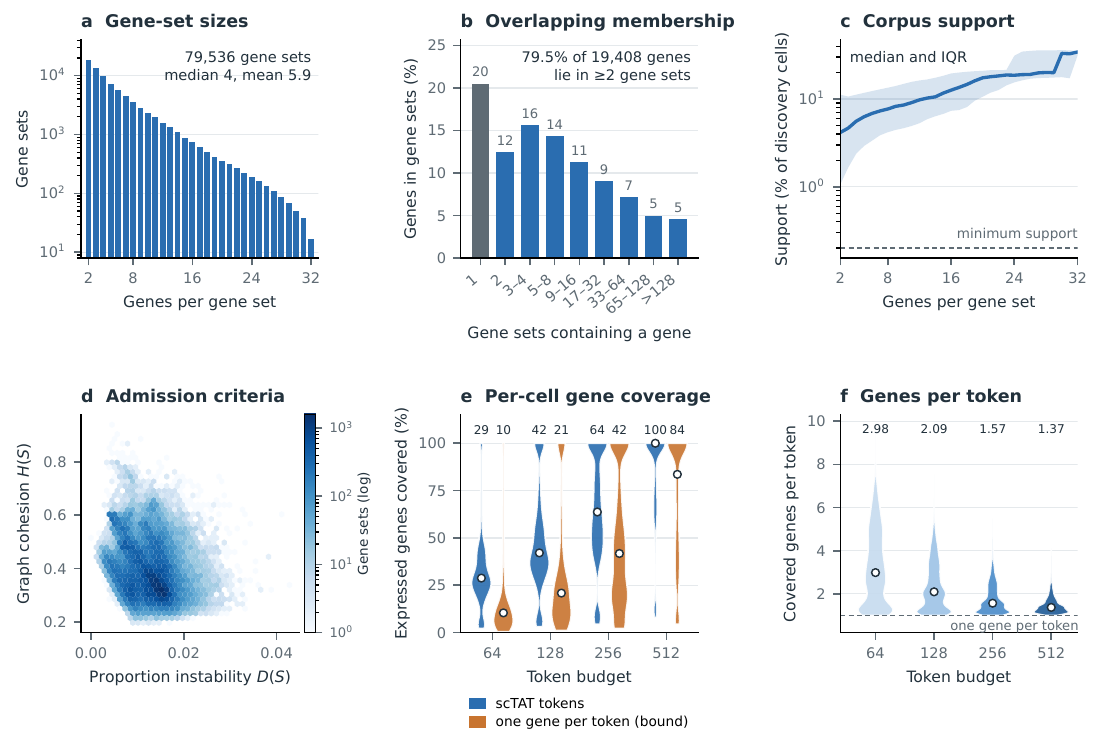}
\caption{Vocabulary statistics (a--d) and per-cell use on 10,000 annotation cells (e--f). \textbf{a}, Gene-set sizes. \textbf{b}, Gene sets containing each gene. \textbf{c}, Support by set size (median and IQR; dashed: admission threshold). \textbf{d}, Proportion instability $D(S)$ versus graph cohesion $H(S)$. \textbf{e}, Expressed genes covered, against the one-gene-per-token bound $\min(K,n)/n$. \textbf{f}, Covered genes per token.}
\label{fig:vocabulary-statistics}
\end{figure}

\paragraph{Vocabulary overview.}
The vocabulary holds 60,694 single genes, 79,536 gene sets, and 3 special tokens (140,233 entries); 19,408 of the 47,495 detectable genes (40.9\%) belong to at least one gene set. Figure~\ref{fig:vocabulary-statistics} summarizes the gene sets and their use. Most sets are small (median 4 genes; panel a), yet they overlap heavily: 79.5\% of the genes in any set belong to at least two (panel b). Support increases with set size but stays above the admission threshold (panel c), and admitted sets combine low proportion instability with moderate graph cohesion (panel d). On 10,000 annotation cells, the selected tokens cover a median of 29\%, 42\%, 64\%, and 100\% of each cell's expressed genes at 64, 128, 256, and 512 tokens, against a median upper bound of 10\%, 21\%, 42\%, and 84\% with one gene per token (panel e); each token then carries a median of 2.98, 2.09, 1.57, and 1.37 genes (panel f). These medians describe the downstream annotation cells, which express fewer genes than the pretraining validation cells summarized by means in Figure~\ref{fig:token-budget}c (median 612.5 and mean 1,507 expressed genes per cell, against a mean of 1,922), so the same budget covers a larger share of each annotation cell.

\paragraph{Enrichment and physical-association definitions.}
We audit the multi-gene entries against pathway membership, curated complexes, and physical associations, independently of the annotation probes. The background comprises 47,495 detectable genes matched by exact symbols, a subset of the full gene vocabulary. Reactome supplies 2,135 eligible pathways~\citep{reactome}; Complex Portal supplies 919 eligible complexes with experimental support~\citep{complexportal}. Enrichment is assessed using upper-tail hypergeometric tests, Bonferroni correction over eligible terms within each entry, and Benjamini--Yekutieli correction across all entries separately for each database, at the prespecified significance level. BioGRID release 5.0.261 supplies 113,621 unique human physical-association pairs supported by at least two distinct PubMed publications~\citep{biogrid}. For an entry of size $m$, association density divides its supported pair count by $\binom{m}{2}$; unmapped members remain in the denominator.

Across all 79,536 multi-gene entries, 10,546 entries (13.26\%) show Reactome enrichment and 2,832 (3.56\%) show enrichment for experimentally supported Complex Portal complexes after within-entry Bonferroni and vocabulary-wide Benjamini--Yekutieli correction.

\paragraph{Null-control construction and testing.}
To reduce redundant entries in the null comparison, we greedily prune entries in descending MDL-score order using a Jaccard-overlap criterion, retaining 55,137 representatives, and uniformly sample a fixed panel of 1,000. The two prespecified endpoints are the fraction of entries with a within-entry Bonferroni-adjusted Reactome hit at the prespecified significance level and the mean multi-publication association density. Three controls globally permute gene labels within strata, preserving entry sizes and all overlaps. The first conditions on detection frequency and annotation availability; the others additionally condition on prior-graph degree, then experimental-network degree. A fourth control draws random stars from the prior graph with the same inferred anchor and size as each entry. The inferred anchor is a deterministically chosen member adjacent to all other members.

Each control uses repeated random draws. One-sided Monte Carlo $P$ values use the plus-one correction; Bonferroni adjustment covers the eight primary comparisons (four controls, two endpoints).

For the fixed panel of 1,000 entries sampled after overlap pruning, the Reactome hit rate is 20.10\% and mean density of physical associations supported by at least two publications is 2.460\%. Here, a pathway hit uses within-entry correction, whereas the vocabulary-wide counts above additionally control across entries. The three stratified permutation controls yield null means of 3.87--4.83\% for pathway hits and 0.153--0.193\% for physical-association density. Conditioning on the same prior-graph anchor and entry size gives higher null means of 17.05\% and 1.778\%, leaving differences of 3.05 and 0.682 percentage points (1.18-fold and 1.38-fold). All eight prespecified one-sided comparisons have Monte Carlo $P=0.005$ and Bonferroni-adjusted $P=0.04$; 0.005 is the resolution limit of the permutation test.

\paragraph{Pathway and complex-supported case studies.}
Figure~\ref{fig:biology-cases} in Section~\ref{sec:biology-results} shows every member of six gene sets. They are the sets selected by the most cells among 2,000 training cells that were not used for vocabulary construction, tokenized at 512 tokens. The three translation-related sets contain 32, 31, and 25 genes, of which 32, 31, and 20 belong to the named Reactome pathways. The two mitochondrial sets contain 11 and 13 genes, of which 11 and 10 map to mitochondrial RNA processing and stability, and both include the MT-CO1/2/3 subunits of complex IV. Their shared annotation reflects regulation of mitochondrial transcripts by FASTK-family proteins~\citep{ohkubo2021}. The sixth set contains 17 genes, including five endosomal/vacuolar pathway members and the HLA-E/B2M complex. The rings separate curated memberships from experimentally recorded physical associations~\citep{reactome,complexportal,biogrid}.

\clearpage

\section{Annotation results with variability}
\label{app:annotation-variability}
\paragraph{Probes and metrics.}
The ten annotation datasets contain 6,595--50,000 cells and 8--25 cell types each, 287,229 cells in total, and are used with their provided labels; vocabulary construction uses only pretraining cells. For annotation, \method{} and the four paired baselines use frozen CLS features, normalized per cell and reused across the five shared train--test splits. Model-specific adapters retain the processed expression values and available gene mappings, so the comparison covers each whole system with its own input adapter. Splits are non-stratified and made at the cell level. KNN uses Euclidean distance and uniform voting. The linear classifier uses shared training settings and the native embedding dimension, and is evaluated once on held-out cells after the prescribed training. Accuracy on a dataset is the fraction of correctly labeled test cells; macro accuracy, the unweighted mean of the ten dataset accuracies, is computed within each split repetition and then averaged.

\paragraph{Variability.}
Tables~\ref{tab:main-knn-std} and~\ref{tab:main-linear-std} provide the full mean $\pm$ sample standard deviation versions of the main annotation tables. They use the same checkpoints, fixed features, input settings, and five shared evaluation splits. For each macro result, the ten datasets are first averaged equally within each repetition; its standard deviation is then computed across the five repetition-wise averages. This variation reflects evaluation splits and, for the linear probe, head initialization, with pretrained weights and features fixed.

For the paired 512-token comparison with expression-weighted scBenchmark, we subtract its macro accuracy from \method{} within each of the five evaluation splits. The mean $\pm$ sample SD of these differences is $0.286\pm0.138$ percentage points for KNN and $0.376\pm0.203$ for the linear probe. Pairing uses the same deterministic split rule on original cell-row order, with matching archived input identities and cell counts.

\begingroup
\makeatletter\setlength{\@fptop}{0pt}\makeatother

\begin{table}[!ht]
\centering
\caption{Frozen KNN annotation accuracy (\%), mean $\pm$ SD over five splits (full version of Table~\ref{tab:main-knn}).}\label{tab:main-knn-std}
\begingroup\setlength{\tabcolsep}{1.3pt}\renewcommand{\arraystretch}{1.12}\footnotesize
\renewcommand{\metricstd}[2]{\begin{tabular}[c]{@{}c@{}}#1\\[-2.5pt]{\fontsize{5.5}{6}\selectfont$\pm$#2}\end{tabular}}
\begin{tabular*}{\linewidth}{@{\extracolsep{\fill}}llc*{10}{c}c@{}}\toprule
\textbf{Model} & \textbf{Input} & \textbf{Budget} & \textbf{MS} & \textbf{Myel.} & \textbf{Panc.} & \textbf{Lupus} & \textbf{Deng.} & \textbf{irCol.} & \textbf{Lepto.} & \textbf{Myas.} & \textbf{Scan.} & \textbf{scF.} & \textbf{Avg} \\
\midrule
\multirow{2}{*}{scBenchmark} & Uniform & \multirow{2}{*}{512} & \metricstd{\ranktwo{85.47}}{0.41} & \metricstd{70.35}{0.48} & \metricstd{96.56}{0.20} & \metricstd{64.93}{0.16} & \metricstd{82.99}{0.32} & \metricstd{53.27}{0.19} & \metricstd{59.54}{0.46} & \metricstd{80.78}{0.21} & \metricstd{92.34}{0.24} & \metricstd{64.46}{0.49} & \metricstd{75.07}{0.17} \\
 & \cellcolor{black!4}Weighted &  & \cellcolor{black!4}\metricstd{\rankone{85.61}}{0.38} & \cellcolor{black!4}\metricstd{71.70}{0.40} & \cellcolor{black!4}\metricstd{97.00}{0.21} & \cellcolor{black!4}\metricstd{64.93}{0.16} & \cellcolor{black!4}\metricstd{\rankone{85.15}}{0.29} & \cellcolor{black!4}\metricstd{\rankone{59.16}}{0.53} & \cellcolor{black!4}\metricstd{90.81}{0.33} & \cellcolor{black!4}\metricstd{\ranktwo{84.33}}{0.26} & \cellcolor{black!4}\metricstd{94.80}{0.11} & \cellcolor{black!4}\metricstd{64.81}{1.00} & \cellcolor{black!4}\metricstd{79.83}{0.16} \\
\midrule[0.3pt]
\multirow{2}{*}{CoCoS} & Uniform & \multirow{2}{*}{512} & \metricstd{81.91}{0.44} & \metricstd{66.69}{0.20} & \metricstd{92.61}{0.34} & \metricstd{\rankone{72.55}}{0.07} & \metricstd{82.68}{0.16} & \metricstd{52.48}{0.44} & \metricstd{61.47}{1.12} & \metricstd{80.63}{0.28} & \metricstd{90.40}{0.37} & \metricstd{\ranktwo{66.48}}{1.01} & \metricstd{74.79}{0.14} \\
 & \cellcolor{black!4}Weighted &  & \cellcolor{black!4}\metricstd{82.15}{0.43} & \cellcolor{black!4}\metricstd{67.77}{0.21} & \cellcolor{black!4}\metricstd{93.04}{0.49} & \cellcolor{black!4}\metricstd{\rankone{72.55}}{0.07} & \cellcolor{black!4}\metricstd{84.53}{0.24} & \cellcolor{black!4}\metricstd{56.18}{0.32} & \cellcolor{black!4}\metricstd{89.37}{0.46} & \cellcolor{black!4}\metricstd{83.65}{0.22} & \cellcolor{black!4}\metricstd{93.34}{0.27} & \cellcolor{black!4}\metricstd{\rankone{66.83}}{1.17} & \cellcolor{black!4}\metricstd{78.94}{0.07} \\
\midrule[0.3pt]
\multirow{2}{*}{scGPT} & Uniform & \multirow{2}{*}{512} & \metricstd{75.42}{0.33} & \metricstd{67.66}{0.45} & \metricstd{91.96}{0.38} & \metricstd{66.49}{0.36} & \metricstd{81.05}{0.52} & \metricstd{49.30}{0.23} & \metricstd{44.67}{0.50} & \metricstd{78.34}{0.38} & \metricstd{90.30}{0.30} & \metricstd{62.99}{0.95} & \metricstd{70.82}{0.10} \\
 & \cellcolor{black!4}Weighted &  & \cellcolor{black!4}\metricstd{75.46}{0.38} & \cellcolor{black!4}\metricstd{68.82}{0.24} & \cellcolor{black!4}\metricstd{92.71}{0.40} & \cellcolor{black!4}\metricstd{66.49}{0.36} & \cellcolor{black!4}\metricstd{84.23}{0.32} & \cellcolor{black!4}\metricstd{55.51}{0.48} & \cellcolor{black!4}\metricstd{87.13}{0.54} & \cellcolor{black!4}\metricstd{83.41}{0.20} & \cellcolor{black!4}\metricstd{93.71}{0.28} & \cellcolor{black!4}\metricstd{63.39}{1.33} & \cellcolor{black!4}\metricstd{77.09}{0.17} \\
\midrule[0.3pt]
\multirow{2}{*}{Geneformer V2} & Uniform & \multirow{2}{*}{512} & \metricstd{70.71}{0.41} & \metricstd{62.90}{0.34} & \metricstd{89.49}{0.28} & \metricstd{\ranktwo{71.98}}{0.26} & \metricstd{76.98}{0.36} & \metricstd{46.58}{0.43} & \metricstd{44.37}{0.77} & \metricstd{74.31}{0.23} & \metricstd{79.66}{0.36} & \metricstd{59.40}{1.03} & \metricstd{67.64}{0.09} \\
 & \cellcolor{black!4}Weighted &  & \cellcolor{black!4}\metricstd{71.28}{0.34} & \cellcolor{black!4}\metricstd{65.35}{0.43} & \cellcolor{black!4}\metricstd{90.16}{0.20} & \cellcolor{black!4}\metricstd{\ranktwo{71.98}}{0.25} & \cellcolor{black!4}\metricstd{78.82}{0.47} & \cellcolor{black!4}\metricstd{50.91}{0.51} & \cellcolor{black!4}\metricstd{72.37}{0.78} & \cellcolor{black!4}\metricstd{78.28}{0.22} & \cellcolor{black!4}\metricstd{87.12}{0.41} & \cellcolor{black!4}\metricstd{60.76}{0.85} & \cellcolor{black!4}\metricstd{72.70}{0.11} \\
\midrule[0.3pt]
\multirow{4}{*}{\method{}} & \multirow{4}{*}{TAT} & 64 & \metricstd{71.89}{0.88} & \metricstd{69.56}{0.38} & \metricstd{\ranktwo{97.57}}{0.17} & \metricstd{66.96}{0.33} & \metricstd{83.49}{0.36} & \metricstd{55.22}{0.17} & \metricstd{92.23}{0.45} & \metricstd{82.46}{0.21} & \metricstd{93.72}{0.29} & \metricstd{62.97}{1.10} & \metricstd{77.61}{0.12} \\
 &  & 128 & \metricstd{76.55}{0.60} & \metricstd{71.86}{0.30} & \metricstd{\rankone{97.68}}{0.29} & \metricstd{68.01}{0.15} & \metricstd{84.17}{0.29} & \metricstd{55.86}{0.51} & \metricstd{93.04}{0.33} & \metricstd{83.47}{0.14} & \metricstd{94.36}{0.25} & \metricstd{64.86}{1.47} & \metricstd{78.99}{0.17} \\
 &  & 256 & \metricstd{80.63}{0.64} & \metricstd{\rankone{72.80}}{0.47} & \metricstd{97.23}{0.36} & \metricstd{68.37}{0.18} & \metricstd{\ranktwo{84.93}}{0.49} & \metricstd{56.76}{0.05} & \metricstd{\rankone{93.27}}{0.47} & \metricstd{84.32}{0.24} & \metricstd{\ranktwo{94.97}}{0.24} & \metricstd{65.89}{0.48} & \metricstd{\ranktwo{79.92}}{0.08} \\
 &  & \cellcolor{black!4}512 & \cellcolor{black!4}\metricstd{81.71}{0.49} & \cellcolor{black!4}\metricstd{\ranktwo{72.75}}{0.29} & \cellcolor{black!4}\metricstd{97.52}{0.19} & \cellcolor{black!4}\metricstd{68.80}{0.28} & \cellcolor{black!4}\metricstd{84.90}{0.22} & \cellcolor{black!4}\metricstd{\ranktwo{57.23}}{0.23} & \cellcolor{black!4}\metricstd{\ranktwo{93.09}}{0.46} & \cellcolor{black!4}\metricstd{\rankone{84.37}}{0.16} & \cellcolor{black!4}\metricstd{\rankone{95.25}}{0.27} & \cellcolor{black!4}\metricstd{65.52}{0.68} & \cellcolor{black!4}\metricstd{\rankone{80.12}}{0.11} \\
\bottomrule\end{tabular*}\endgroup
\end{table}

\clearpage
\begin{table}[!ht]
\centering
\caption{Frozen linear-probe annotation accuracy (\%), mean $\pm$ SD over five splits (full version of Table~\ref{tab:main-linear}).}\label{tab:main-linear-std}
\begingroup\setlength{\tabcolsep}{1.3pt}\renewcommand{\arraystretch}{1.12}\footnotesize
\renewcommand{\metricstd}[2]{\begin{tabular}[c]{@{}c@{}}#1\\[-2.5pt]{\fontsize{5.5}{6}\selectfont$\pm$#2}\end{tabular}}
\begin{tabular*}{\linewidth}{@{\extracolsep{\fill}}llc*{10}{c}c@{}}\toprule
\textbf{Model} & \textbf{Input} & \textbf{Budget} & \textbf{MS} & \textbf{Myel.} & \textbf{Panc.} & \textbf{Lupus} & \textbf{Deng.} & \textbf{irCol.} & \textbf{Lepto.} & \textbf{Myas.} & \textbf{Scan.} & \textbf{scF.} & \textbf{Avg} \\
\midrule
\multirow{2}{*}{scBenchmark} & Uniform & \multirow{2}{*}{512} & \metricstd{\rankone{87.18}}{0.52} & \metricstd{69.74}{0.44} & \metricstd{96.07}{0.20} & \metricstd{76.67}{0.19} & \metricstd{82.44}{0.45} & \metricstd{52.29}{0.70} & \metricstd{66.47}{0.58} & \metricstd{81.96}{0.24} & \metricstd{92.98}{0.36} & \metricstd{67.31}{1.50} & \metricstd{77.31}{0.24} \\
 & \cellcolor{black!4}Weighted &  & \cellcolor{black!4}\metricstd{\ranktwo{87.09}}{0.56} & \cellcolor{black!4}\metricstd{70.58}{0.55} & \cellcolor{black!4}\metricstd{96.74}{0.18} & \cellcolor{black!4}\metricstd{76.67}{0.19} & \cellcolor{black!4}\metricstd{84.70}{0.52} & \cellcolor{black!4}\metricstd{56.16}{0.47} & \cellcolor{black!4}\metricstd{91.63}{0.38} & \cellcolor{black!4}\metricstd{85.00}{0.31} & \cellcolor{black!4}\metricstd{94.60}{0.40} & \cellcolor{black!4}\metricstd{67.55}{1.53} & \cellcolor{black!4}\metricstd{81.07}{0.26} \\
\midrule[0.3pt]
\multirow{2}{*}{CoCoS} & Uniform & \multirow{2}{*}{512} & \metricstd{86.12}{0.47} & \metricstd{68.28}{0.61} & \metricstd{95.00}{0.15} & \metricstd{78.03}{0.15} & \metricstd{83.63}{0.48} & \metricstd{53.02}{0.51} & \metricstd{68.96}{0.45} & \metricstd{81.96}{0.28} & \metricstd{92.40}{0.45} & \metricstd{\rankone{69.38}}{1.02} & \metricstd{77.68}{0.15} \\
 & \cellcolor{black!4}Weighted &  & \cellcolor{black!4}\metricstd{86.10}{0.41} & \cellcolor{black!4}\metricstd{68.65}{0.75} & \cellcolor{black!4}\metricstd{95.45}{0.23} & \cellcolor{black!4}\metricstd{78.03}{0.15} & \cellcolor{black!4}\metricstd{\ranktwo{85.53}}{0.52} & \cellcolor{black!4}\metricstd{\ranktwo{56.22}}{0.30} & \cellcolor{black!4}\metricstd{91.13}{0.18} & \cellcolor{black!4}\metricstd{84.82}{0.22} & \cellcolor{black!4}\metricstd{94.27}{0.17} & \cellcolor{black!4}\metricstd{\ranktwo{69.29}}{0.92} & \cellcolor{black!4}\metricstd{80.95}{0.11} \\
\midrule[0.3pt]
\multirow{2}{*}{scGPT} & Uniform & \multirow{2}{*}{512} & \metricstd{81.71}{0.16} & \metricstd{70.28}{0.47} & \metricstd{93.78}{0.20} & \metricstd{76.66}{0.12} & \metricstd{83.80}{0.50} & \metricstd{52.39}{0.56} & \metricstd{56.43}{0.48} & \metricstd{81.92}{0.30} & \metricstd{92.38}{0.29} & \metricstd{68.52}{1.17} & \metricstd{75.79}{0.15} \\
 & \cellcolor{black!4}Weighted &  & \cellcolor{black!4}\metricstd{81.55}{0.21} & \cellcolor{black!4}\metricstd{71.24}{0.56} & \cellcolor{black!4}\metricstd{94.39}{0.13} & \cellcolor{black!4}\metricstd{76.66}{0.12} & \cellcolor{black!4}\metricstd{\rankone{86.13}}{0.54} & \cellcolor{black!4}\metricstd{\rankone{57.33}}{0.52} & \cellcolor{black!4}\metricstd{90.46}{0.35} & \cellcolor{black!4}\metricstd{\rankone{85.60}}{0.24} & \cellcolor{black!4}\metricstd{\ranktwo{95.23}}{0.23} & \cellcolor{black!4}\metricstd{68.52}{1.11} & \cellcolor{black!4}\metricstd{80.71}{0.12} \\
\midrule[0.3pt]
\multirow{2}{*}{Geneformer V2} & Uniform & \multirow{2}{*}{512} & \metricstd{80.81}{0.51} & \metricstd{65.90}{0.63} & \metricstd{92.87}{0.27} & \metricstd{\rankone{80.02}}{0.22} & \metricstd{80.30}{0.40} & \metricstd{50.72}{0.59} & \metricstd{50.32}{0.81} & \metricstd{79.95}{0.44} & \metricstd{90.92}{0.41} & \metricstd{63.79}{1.17} & \metricstd{73.56}{0.21} \\
 & \cellcolor{black!4}Weighted &  & \cellcolor{black!4}\metricstd{81.29}{0.38} & \cellcolor{black!4}\metricstd{67.03}{0.81} & \cellcolor{black!4}\metricstd{94.00}{0.22} & \cellcolor{black!4}\metricstd{\rankone{80.02}}{0.22} & \cellcolor{black!4}\metricstd{81.65}{0.61} & \cellcolor{black!4}\metricstd{54.04}{0.63} & \cellcolor{black!4}\metricstd{83.12}{0.69} & \cellcolor{black!4}\metricstd{82.65}{0.31} & \cellcolor{black!4}\metricstd{93.98}{0.30} & \cellcolor{black!4}\metricstd{64.16}{1.15} & \cellcolor{black!4}\metricstd{78.19}{0.18} \\
\midrule[0.3pt]
\multirow{4}{*}{\method{}} & \multirow{4}{*}{TAT} & 64 & \metricstd{78.76}{0.46} & \metricstd{69.47}{0.74} & \metricstd{\rankone{97.54}}{0.20} & \metricstd{76.64}{0.27} & \metricstd{83.00}{0.50} & \metricstd{54.20}{0.57} & \metricstd{92.50}{0.32} & \metricstd{83.94}{0.33} & \metricstd{94.23}{0.27} & \metricstd{66.93}{0.75} & \metricstd{79.72}{0.12} \\
 &  & 128 & \metricstd{81.67}{0.55} & \metricstd{70.82}{0.41} & \metricstd{97.33}{0.28} & \metricstd{77.08}{0.15} & \metricstd{84.07}{0.42} & \metricstd{54.54}{0.66} & \metricstd{92.87}{0.16} & \metricstd{84.77}{0.29} & \metricstd{94.62}{0.21} & \metricstd{67.50}{0.97} & \metricstd{80.53}{0.19} \\
 &  & 256 & \metricstd{85.13}{0.45} & \metricstd{\rankone{71.56}}{0.24} & \metricstd{96.96}{0.38} & \metricstd{77.59}{0.32} & \metricstd{84.43}{0.51} & \metricstd{54.84}{0.74} & \metricstd{\ranktwo{93.07}}{0.26} & \metricstd{85.25}{0.33} & \metricstd{94.93}{0.33} & \metricstd{68.37}{1.45} & \metricstd{\ranktwo{81.21}}{0.28} \\
 &  & \cellcolor{black!4}512 & \cellcolor{black!4}\metricstd{85.74}{0.32} & \cellcolor{black!4}\metricstd{\ranktwo{71.31}}{0.24} & \cellcolor{black!4}\metricstd{\ranktwo{97.43}}{0.28} & \cellcolor{black!4}\metricstd{\ranktwo{78.11}}{0.20} & \cellcolor{black!4}\metricstd{84.18}{0.39} & \cellcolor{black!4}\metricstd{55.28}{0.56} & \cellcolor{black!4}\metricstd{\rankone{93.25}}{0.41} & \cellcolor{black!4}\metricstd{\ranktwo{85.26}}{0.18} & \cellcolor{black!4}\metricstd{\rankone{95.28}}{0.34} & \cellcolor{black!4}\metricstd{68.65}{1.13} & \cellcolor{black!4}\metricstd{\rankone{81.45}}{0.16} \\
\bottomrule\end{tabular*}\endgroup
\end{table}

\FloatBarrier
\endgroup

\section{Gene-Network Evaluation Details}
\label{app:grn-evaluation}
\paragraph{Input and scoring.}
Following the Han hESC workflow of scPRINT~\citep{scprint}, all models start from the same panel: the first 1,024 Han hESC cells and 5,000 highly variable genes. Each model keeps its native tokenizer, preprocessing, and vocabulary filtering, which leaves 4,984--5,000 scorable genes. BenGRN~1.2.4 scores all candidate edges against the Full, ChIP, and KO references (22, 41, and 112--113 transcription factors), and we report trapezoidal AUPRC without a held-out edge split or label-based head selection. The random baseline is the positive-edge prevalence of each reference, which varies by less than 0.001 across models; Table~\ref{tab:grn} reports its mean.

\paragraph{Models and extraction.}
GENIE3~\citep{genie3} is fitted on the expression panel; all pretrained encoders are frozen. scGPT~\citep{scgpt} uses its whole-human checkpoint and the reference workflow's last-layer, eight-head mean. Geneformer~\citep{geneformer} uses the 12L--30M checkpoint with its native ranked input and the mean over all layers and heads. scPRINT uses the arithmetic mean of its 128 heads. \method{} (the main 512-token model) averages its 48 heads and pools gene-level query/key representations from its native input, with zero scores for genes it does not cover. Query--key scores are transposed once to the scorer's source--target orientation.

\section{Perturbation Evaluation Details}
\label{app:perturbation-evaluation}
\paragraph{Shared data and splits.}
We follow the Norman\_1 workflow in PertEval-scFM~\citep{perteval}, starting from its original Norman raw-data source. After eligibility filtering, the task contains 11,855 control cells, 56,147 treated cells, 101 perturbation conditions, and 2,060 target genes. Eligibility requires usable controls expressing the perturbed gene. One condition without such controls is excluded under the upstream rule. All methods in Table~\ref{tab:perturbation}(b) use the same 101 retained conditions, 2,060 target genes, test cells, saved SPECTRA splits, and control-pooling draws. Gene identities are resolved without summing duplicate-symbol counts; encoder-input name mapping preserves the original target columns and perturbation labels. We use the official split generator.

\paragraph{Control-derived features.}
For pretrained models, basal features average embeddings over shared draws of control cells. A condition feature averages embeddings of eligible controls after removing or nullifying the perturbed gene. The predictor receives the concatenated basal and condition features. Observed treated expression supplies supervised targets and SPECTRA split construction, but never encoder features.

\method{} uses frozen, unnormalized CLS embeddings from the main 512-token model, which also tokenizes its inputs. Its benchmark adapter normalizes and log-transforms control expression before nullifying the target gene. Geneformer uses the six-layer, 256-dimensional 6L--30M checkpoint specified by PertEval-scFM, whereas annotation uses its V2 checkpoint. It tokenizes raw controls by median-normalized rank, removes the target gene, and averages the non-padding final-layer embeddings. The scFoundation model~\citep{scfoundation} uses its native single-cell adapter with the target count zeroed before normalization, and concatenates its two special-token readouts with maximum and mean gene embeddings. scGPT uses the whole-human checkpoint, native positive-gene binning and sampling, and per-cell L2-normalized CLS embeddings before control pooling. No additional L2 normalization is applied after pooling. Each model thus uses its own adapter and readout.

The raw-expression MLP uses control-derived expression pseudo-bulks and condition features given by the perturbed gene's Pearson correlations with target genes, following the upstream reference. ``Raw-expression'' here denotes the absence of a pretrained encoder, not unprocessed sequencing counts. The Mean baseline preserves the upstream batch-dependent rule: it predicts each test batch's mean basal expression minus the corresponding basal vector, with no learned predictor.

\paragraph{Targets, fitting, and scoring.}
Normalization, target-gene selection, and SPECTRA split construction follow the upstream transductive preparation using all retained data. Targets measure expression changes in normalized/log space, retaining the upstream subset-local basal subtraction. Test conditions are held out from predictor fitting, while validation cells are drawn from training conditions. MLPs share the predictor recipe, with widths adapted to the representation dimension, and are selected by validation MSE before one final test evaluation.

Test MSE weights all retained cell--gene elements equally. Table~\ref{tab:perturbation}(b) reports means and sample standard deviations across three paired split/predictor repetitions with fixed encoder weights. Each integrated score is computed within a repetition by trapezoidal integration over the stated $s$ interval and is not normalized by interval length.
\paragraph{Additional tasks.}
Table~\ref{tab:perturbation-additional} extends the comparison to the Norman double-gene, Replogle K562, and Replogle RPE1 tasks of PertEval-scFM, using the same features and predictor for \method{}. \method{} is comparable to the other pretrained encoders on all three.

\begin{table}[!htbp]
\centering
\caption{Additional PertEval-scFM tasks: area under the test-MSE curve ($\times10^{-2}$, lower is better) over $s\in[0,0.7]$ for Norman and $s\in[0.1,0.7]$ for Replogle. Baseline values are as reported by PertEval-scFM~\citep{perteval}; \method{} is evaluated here (mean $\pm$ SD over three repetitions).}\label{tab:perturbation-additional}
\begingroup\small\setlength{\tabcolsep}{5pt}\renewcommand{\arraystretch}{1.12}
\renewcommand{\metricstd}[2]{\mbox{#1\,{\fontsize{6}{7}\selectfont$\pm$#2}}}
\begin{tabular*}{0.8\linewidth}{@{\extracolsep{\fill}}lccc@{}}
\toprule
\textbf{Model} & \textbf{Norman double} & \textbf{Replogle K562} & \textbf{Replogle RPE1} \\
\midrule
Mean & \metricstd{4.26}{0.07} & \metricstd{13.86}{0.76} & \metricstd{13.41}{0.35} \\
Raw-expr.\ MLP & \metricstd{4.25}{0.07} & \metricstd{14.20}{0.77} & \metricstd{12.52}{0.24} \\
Geneformer & \metricstd{4.50}{0.08} & \metricstd{14.75}{0.56} & \metricstd{12.51}{0.23} \\
scBERT & \metricstd{4.53}{0.08} & \metricstd{12.28}{0.16} & \metricstd{12.58}{0.27} \\
scFoundation & \metricstd{4.43}{0.47} & \metricstd{15.56}{1.13} & \metricstd{12.53}{0.23} \\
scGPT & \metricstd{5.18}{0.13} & \metricstd{13.84}{0.80} & \metricstd{12.53}{0.23} \\
UCE & \metricstd{4.44}{0.08} & \metricstd{13.84}{0.80} & \metricstd{12.53}{0.24} \\
\midrule[0.3pt]
\method{} & \metricstd{4.41}{0.40} & \metricstd{13.84}{0.55} & \metricstd{12.55}{0.02} \\
\bottomrule
\end{tabular*}
\endgroup
\end{table}

\section{Vocabulary Construction and Settings}
\label{app:vocabulary-construction}
\paragraph{Occurrence costs.}
Let $N$ be the number of discovery cells, $G=|\mathcal G|$, $n_g$ the number of cells in which gene $g$ is active, and $n_S$ the number in which every member of $S$ is active. For the fixed positive pseudocount $\kappa$, the score uses
\begin{equation}
\widetilde p_g=\frac{n_g+\kappa}{N+2\kappa},\qquad
\widetilde p_S=\frac{n_S+\kappa}{N+2\kappa},\qquad
\ell_g=-\log_2\widetilde p_g,\quad
\ell_S=-\log_2\widetilde p_S.
\end{equation}
These are smoothed probabilities of separate presence events ($\kappa=0.5$), held fixed during search and not normalized across genes or entries. Let $B_{\mathrm{size}}$ be the number of size codes in the fixed multi-gene cardinality codebook. The description penalty is
\begin{equation}
L_{\mathrm{desc}}(S)=\log_2 B_{\mathrm{size}}+
\log_2\binom{G}{|S|}.
\end{equation}
It charges for entry size and its unordered member set. This comparison follows description-length pattern selection~\citep{krimp} and is applied to each candidate separately.

\paragraph{Expression and prior penalties.}
For each supporting cell, define $p_{ig}=x_{ig}/\sum_{h\in S}x_{ih}$ for $g\in S$, and $\bar p_S=n_S^{-1}\sum_{i\in I_S}p_i$. The expression penalty is
\begin{equation}
D(S)=\frac{1}{n_S}\sum_{i\in I_S}\operatorname{JSD}_2(p_i,\bar p_S),
\quad
\operatorname{JSD}_2(p,q)=\tfrac12\operatorname{KL}_2(p\|m)+\tfrac12\operatorname{KL}_2(q\|m),
\quad m=\tfrac12(p+q).
\end{equation}
For a maximum spanning tree $T_S$ of the prior subgraph on $S$, let
\begin{equation}
H(S)=\frac{1}{|S|-1}\sum_{(g,h)\in T_S}w_{gh},
\qquad \mathcal R(S)=D(S)+\frac{1-H(S)}{2}.
\end{equation}
Base-two Jensen--Shannon divergence lies in $[0,1]$, and cosine edge weights lie in $[-1,1]$. The two penalty components thus have the same bounded range and are added with unit coefficients before multiplication by support. This scaling defines the admission heuristic.

\paragraph{Search procedure.}
Algorithm~\ref{alg:vocab} gives the offline search with the admission score $F$ of Equation~\ref{eq:gain}, minimum support $\tau$ in each of two disjoint discovery subsets, and maximum set size $m_{\max}$. Ties are broken by set size, graph coherence, and then a fixed lexicographic order. An upper bound obtained by omitting the nonnegative expression and prior penalties prunes extensions without changing the selected increment. Because the same observation may support several overlapping entries, the scores of different entries are not additive. The vocabulary is frozen before encoder training.

\begin{algorithm}[!ht]
\begin{minipage}{\linewidth}
\setlength{\abovecaptionskip}{3pt}\setlength{\belowcaptionskip}{3pt}
\hrule height .8pt
\caption{Vocabulary construction (offline)}\label{alg:vocab}
\hrule\vspace{4pt}
{\footnotesize\noindent\algkw{Input:} graph neighbors $\mathcal N(a)$; discovery subsets $D_1,D_2$; support $\tau$; maximum size $m_{\max}$\par\vspace{2pt}}
\begin{algsteps}
\algstep{0}{$\mathcal C\gets\varnothing$}
\algstep{0}{\algkw{for} each anchor gene $a\in\mathcal G$ \algkw{do}}
\algstep{1}{$S\gets\{a\}$;\ \ $f\gets 0$}
\algstep{1}{\algkw{while} $|S|<m_{\max}$ \algkw{do}}
\algstep{2}{$\mathcal E\gets\{S\cup\{g\}:g\in\mathcal N(a)\setminus S\}$, keeping sets active in at least $\tau$ cells of both $D_1$ and $D_2$}
\algstep{2}{\algkw{if} $\mathcal E=\varnothing$ or $\max_{S'\in\mathcal E}F(S')\le f$ \algkw{then break}}
\algstep{2}{$S\gets\arg\max_{S'\in\mathcal E}F(S')$}
\algstep{2}{$f\gets F(S)$;\ \ $\mathcal C\gets\mathcal C\cup\{S\}$}
\algstep{0}{Drop duplicates and strict subsets with the same supporting cells}
\algstep{0}{\algkw{return} $\mathcal V=\mathcal C\cup\{\{g\}:g\in\mathcal G\}$}
\end{algsteps}
\vspace{2pt}\hrule height .8pt
\end{minipage}
\end{algorithm}

\paragraph{Control vocabularies.}
The \textit{w/o MDL} vocabulary (Table~\ref{tab:ablation-vocab}b) keeps the graph and the support requirement but drops the admission score. Each candidate grows from an anchor gene by adding its most similar graph neighbors while the set remains co-detected in at least 50 cells in each of the two discovery subsets; for each set size, we keep, where available, as many of the most graph-coherent candidates as the main vocabulary contains. Only 2,031 two-gene sets could not be matched, giving 77,505 gene sets against 79,536. The two vocabularies thus share the graph, the support requirement, and nearly the same size distribution, and differ only in whether the admission score decides which sets are kept. The \textit{w/o gene content} vocabulary permutes gene labels within 32 detection-frequency strata, preserving the number, sizes, and overlaps of entries while discarding their gene content. Both controls are pretrained at 512 tokens, \textit{w/o MDL} with the \method{} tokenizer and \textit{w/o gene content} with Expression-first selection.

\paragraph{Settings.}
Table~\ref{tab:config} lists the settings of the vocabulary and of the main \method{} models; the control vocabularies reuse the pretraining settings.
\begin{table}[!ht]
\centering
\caption{Settings of the vocabulary and the main \method{} models.}\label{tab:config}
\begingroup\footnotesize\setlength{\tabcolsep}{3pt}\renewcommand{\arraystretch}{1.1}
\begin{tabular*}{\linewidth}{@{\extracolsep{\fill}}llll@{}}
\toprule
\textbf{Setting} & \textbf{Value} & \textbf{Setting} & \textbf{Value} \\
\midrule
Gene graph & mutual 32-NN on scGPT genes & Graph size & 60,694 genes, 202,582 edges \\
Discovery sample & $2\times25{,}000$ cells & Min.\ support $\tau$ & 50 cells per subset \\
Maximum set size $m_{\max}$ & 32 genes & Pseudocount $\kappa$ & 0.5 \\
Vocabulary & 79,536 gene sets, 60,694 genes & Parameters & 22.2M \\
\midrule
Encoder & 6 layers, width 256, 8 heads & FFN; dropout & 256; 0.15 \\
Pretraining data & 1.8M cells, 10 epochs & Batch; updates & 256 cells; 70,320 \\
Objective & MLM + MVC, equal weights & Mask ratio & 0.4 \\
Optimizer & Adam, lr $2\times10^{-4}$, clip 1.0 & Schedule & 10k-update warmup, constant \\
Precision & BF16, FP32 loss & Seed & 42 \\
\bottomrule
\end{tabular*}\endgroup
\end{table}

\end{document}